# An integrated system built for small-molecule semiconductors via high-throughput approaches

Jianchang Wu[1,2]*, Jiyun Zhang[1,2], Manman Hu[3], Patrick Reiser[4], Luca Torresi[4], Pascal Friederich[4,5], Leopold Lahn[2,6], Olga Kasian[2,6], Dirk M. Guldi[7], M. Eugenia Pérez-Ojeda[8], Anastasia Barabash[1,2], Juan S. Rocha-Ortiz[2], Yicheng Zhao[1,2,9], Zhiqiang Xie[2], Junsheng Luo[2,9], Yunuo Wang[2], Sang Il Seok[3], Jens A. Hauch[1,2], and Christoph J. Brabec[1,2]*

[1]Forschungszentrum Jülich GmbH, Helmholtz-Institute Erlangen-Nürnberg (HI-ERN), Immerwahrstraße 2, 91058 Erlangen, Germany

[2]Friedrich-Alexander-Universität Erlangen-Nürnberg (FAU), Faculty of Engineering, Department of Material Science, Materials for Electronics and Energy Technology (i-MEET), Martensstrasse 7, 91058 Erlangen, Germany

[3]Department of Energy Engineering, School of Energy and Chemical Engineering, Ulsan National Institute of Science and Technology (UNIST), 50 UNIST-gil, Eonyang-eup, Ulju-gun, Ulsan 44919, Korea

[4]Institute of Nanotechnology, Karlsruhe Institute of Technology (KIT), Hermann-von-Helmholtz-Platz 1, 76344 Eggenstein-Leopoldshafen, Germany

[5]Institute of Theoretical Informatics, Karlsruhe Institute of Technology (KIT), Ham Fasanengarten 5, 76131 Karlsruhe, Germany

[6]Helmholtz-Zentrum Berlin GmbH, Helmholtz Institut Erlangen-Nürnberg, Cauerstraße 1, 91058 Erlangen, Germany

[7]Department of Chemistry and Pharmacy & Interdisciplinary Center of Molecular Materials (ICMM), Friedrich-Alexander-Universität Erlangen-Nürnberg (FAU), Erlangen 91058, Germany

[8]Department of Chemistry and Pharmacy, Friedrich-Alexander-Universität Erlangen-Nürnberg (FAU), Nikolaus-Fiebiger-Straße 10, 91058 Erlangen, Germany

[9]University of Electronic Science and Technology of China, School of Electronic Science and Engineering, State Key Laboratory of Electronic Thin Films and Integrated Devices, 611731 Chengdu, P. R. China.

Corresponding Authors

E-mail: jianchang.wu@fau.de; christoph.brabec@fau.de

## Abstract

High-throughput synthesis of solution-processable structurally variable small-molecule semiconductors is both an opportunity and a challenge. A large number of diverse molecules provide a possibility for quick material discovery and machine learning based on experimental data. However, the diversity of molecular structure leads to the complexity of molecular properties, such as solubility, polarity, and crystallinity, which poses great challenges to solution processing and purification. Here, we first report an integrated system for the high-throughput synthesis, purification, and characterization of molecules with a large variety. Based on the principle 'Like dissolves like', we combine theoretical calculations and a robotic platform to accelerate the purification of those molecules. With this platform, a material library containing 125 molecules and their optical-electric properties was built within a timeframe of weeks. More importantly, the high repeatability of recrystallization we design is a reliable approach to further upgrading and industrial production.

**INTRODUCTION**

Conjugated small molecules have been used in a vast number of optoelectronic applications, such as organic light-emitting diodes (OLEDs)[1], organic solar cells (OSCs)[2-3], perovskite solar cells (PSCs)[4-5], organic thin-film transistors (OTFTs)[6], and optical imaging[7] owing to their unique optical-electronic properties, batch-to-batch reproducibility, and solution-processability. Breakthroughs in materials science always rely on two key factors, namely establishing broadly applicable quantitative structure-property relationships and discovering materials beyond current rules. Implicit in the former is, for example, the relationship between open circuit voltage ($V_{oc}$) and the energy difference between the highest occupied molecular orbital (HOMO) of the electron donor and the lowest unoccupied molecular orbital (LUMO) of the electron acceptor[8-9]. This per se provides the direction for the design of conjugated molecules and is essential for promoting the performance of organic solar cells beyond 10%. The latter has opened up in the form of conducting polymers' new realms for investigation[10-11]. Achieving these two goals calls for a material library containing solution-processable structurally variable small-molecule semiconductors and the corresponding properties. Literally, millions of molecules could be synthesized via the coupling of different commercial building blocks. Moreover, their optoelectronic properties are tunable by means of employing proper building blocks. This suggests that creating a material library containing a large number of diverse molecules is feasible. With this material library at hand, we should be able to construct more widely applicable structure-property relationships. These enable not only the rational design and synthesis of tailor-made materials but also the potential discovery of materials with unexpected properties. However, the individual synthetic procedure, purification, and properties evaluation process of a large number of materials are time-consuming and labor-intense.

Solid-phase synthesis offers an efficient way to realize high-throughput (HT) organic syntheses[12]. Bäuerle *et al.* prepared, for example, a library of 256 oligo(3-arylthiophene)s by utilizing solid-phase synthesis[13-14]. It is based on the stepwise addition of functional building blocks to a growing conjugated oligothiophene chain, which was covalently bound to a solid resin particle. It provided a procedure whereby reagents and by-products were simply removed by filtration, and recrystallization of any intermediates was eliminated. Other types of π-conjugated oligomers have been reported and synthesized by a solid-phase approach, including oligo-(dialkylfluorene)s[15], oligo-(triarylamine)s[16], and alternating co-oligomers thereof[17]. Also notable is that Burke *et al.* developed a new type of catch-and-release purification protocol for N-methyliminodiacetic acid (MIDA)-boronate-containing intermediates[18-19]. The MIDA boronates absorbed on silica gel were purified by switching the eluent, thanks to their binary elution properties. It reflects a general and automated purification process based on automated iterative MIDA-boronates assembly. However, both solid-phase syntheses and MIDA boronates based platforms serve to simplify the purification of the intermediates in the multi-step synthesis. Still, the final products require to be processed and purified by conventional means, which are unsuitable for one-step reactions. Meanwhile, starting from the commercial building blocks, one-step coupling reactions afford up to millions of molecules with different properties that are promising for semiconductor device applications.

Thus oriented, here, we present a semi-automatic platform to conduct the HT one-step synthesis and purification by combining a microwave reactor, vacuum manifold, sample concentrator, and robot system. Based on the principle 'Like dissolves like', we combine theoretical calculations and a robotic platform to accelerate the purification of those molecules. With this platform, more than 20 products were purified in parallel. In total, more than 100 small-molecule organic semiconductors with large variations in molecular structure have been synthesized, purified, and characterized. All of them were analyzed by $^1$H NMR to confirm their purity. Their optoelectronic properties were characterized by means of absorption, photoluminescence, mobility, and cyclic voltammetry using HT characterization equipment.

**RESULTS AND DISCUSSION**

The entire workflow consists of three high-throughput (HT) processes: HT synthesis, HT purification, and HT characterization. In the first step, the reactions are carried out in a microwave reactor. This allows to shorten the reaction time from 1 day to one hour and also runs automatically up to 48 reactions one by one. After full optimization, the Suzuki-Miyaura coupling reaction is completed within 30 min with over 95% yield. This per se reduces the difficulty of subsequent purification greatly. For the second step of purification, we developed a two-step purification process that combines filtration and recrystallization. Filtration, on the one hand, helps to remove the reagents from the reaction mixture, such as the catalyst, acidic raw materials, and inorganic salts. Recrystallization, on the other hand, is used for further purification.

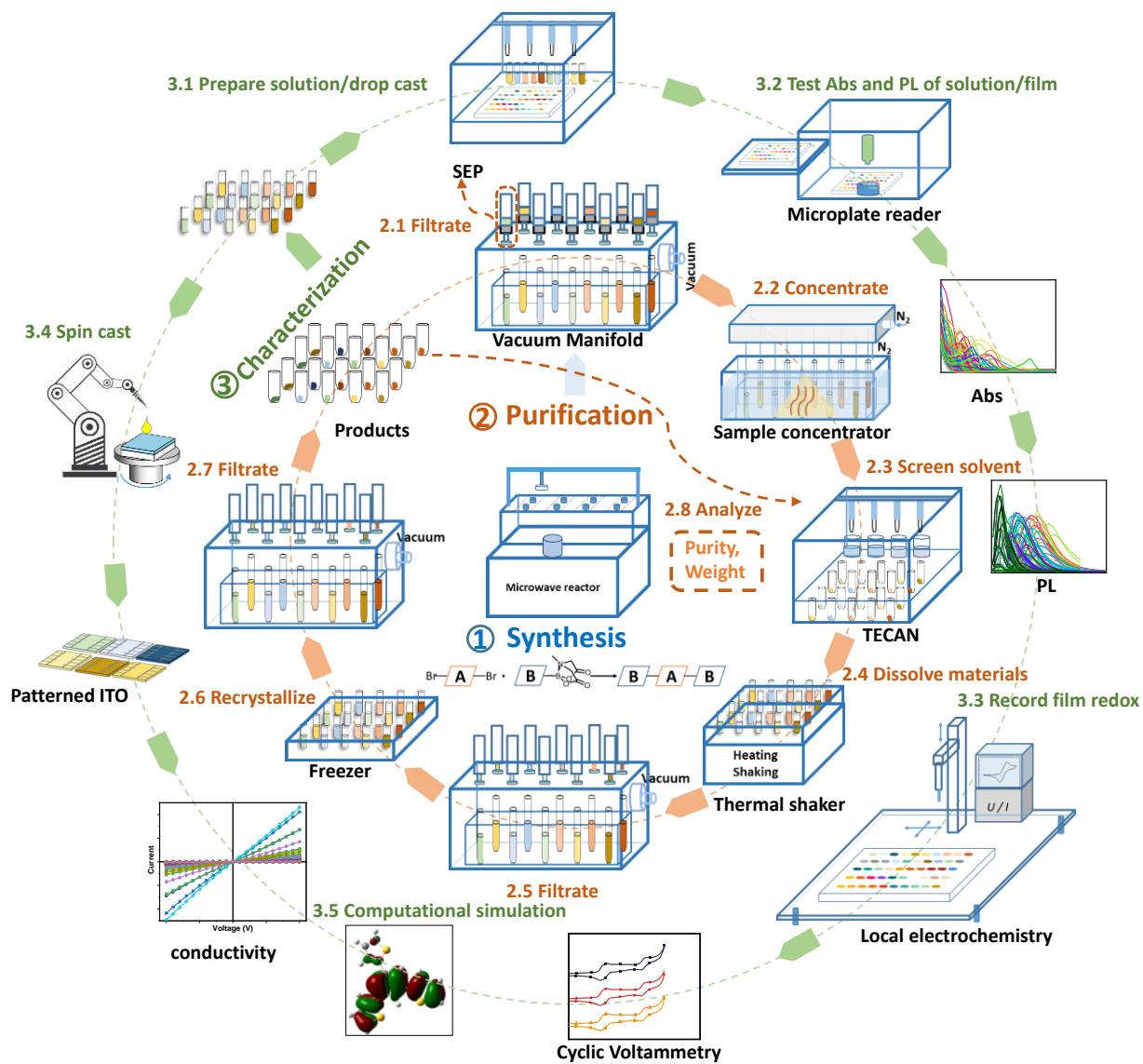

**Figure 1. Workflow of the high-throughput synthesis, purification, and characterization.** 1. HT synthesis with microwave reactor; 2. HT purification: filtrate, concentrate, screen solvent, dissolve materials, filtrate, recrystallize, filtrate, and analyze; 3. HT characterization: prepare solution/drop cast, test UV-vis absorption and PL in solution/film, record films redox, spin cast/mobility test, computational simulation.

In detail, the organic layers of the reaction mixtures are transferred into the vacuum manifold, which speeds up the filtration process. They are then filtered through a homemade solid phase extraction (SEP) (step 2.1), which consists of activated carbon, basic alumina, and silica gel, followed by solvent removal using a sample concentrator (step 2.2). Then, the recrystallization is implemented through solvent screening by a liquid handling and automation system (TECAN, step 2.3), dissolving material (step 2.4), filtrating (step 2.5, removing insoluble impurities by cotton), and freezing (step 2.6).

After freezing for a few hours, crystals precipitate from the solution and are collected by removing the solvent through filtration (step 2.7, filter: cotton). Throughout this process, some materials may be in the form of crystals from recrystallization (step 2.7) or filter cake from filtration (steps 2.1 and 2.5), especially for materials with low solubility. We name them product-C and product-F, respectively. Sample purity and sample weight are decisive in terms of another recrystallization step, which will be discussed in detail in the next part (HT synthesis and purification). The purified materials are characterized by means of absorption, photoluminescence (PL), cyclic voltammetry (CV), and mobility. Absorption and PL are tested both in solution and thin films. We use an in-house developed HT electrochemical setup based on local electrochemistry for the CV characterization of films formed by drop casting through the TECAN robotic system. In addition, an in-house spin coating robot allows depositing thin films on the patterned ITO substrates and measuring the mobility. Theoretical properties, including molecular dipole, energy, and energy levels, are obtained by computational simulation.

**Reaction optimization based on reference materials Re1 and Re2.** We used a microwave-assisted Suzuki coupling reaction to synthesize the material library. To first optimize one cycle of HT synthesis, HT purification, and HT characterization, we subjected methoxy-substituted triphenylamines MIDA ester (**B1**) and phenyl MIDA ester (**B7**) to Suzuki coupling with 2,5-dibromothiophene (**A34**) and tris(4-bromophenyl)amine (**A30**), respectively. We termed the products **Re1** (2,5-Bis(4,4'-bis(methoxyphenyl)aminophen-4''-yl)-thiophene) and **Re2** (Tris(4-biphenylyl)amine), respectively (Figure 2d). Here, we mainly focus on p-type organic semiconductors, which is the case for most **Re1** and **Re2**. Differences in the structure of the two reference materials allow determining a general approach to process most molecules of the material library.

For the HT monitoring of the yield of the reactions, we established the corresponding concentration-absorbance dependencies based on **Re1**[20] and **Re2** (Figure 2a and Supplementary Figure 1) before the syntheses. The **Re1** and **Re2** yields under various reaction conditions were calculated from the ratio of the absorbances at 404 nm for **Re1** or 344 nm for **Re2** of the diluted solutions to the theoretical value of 100% yield (Figure 2b and 2c). It is worth mentioning that the remaining non-reacted starting materials have almost no absorption after 300 nm and thus no influence on the yield calculation.

All absorption data were obtained by HT Microplate Reader. To obtain the optimal reaction protocol, the reaction time, the temperature, the solvent, the catalyst, the base, and the stoichiometric ratio (equiv.) of the reaction were systematically optimized. We found that the equiv. of boronic monomer relative to bromine has a significant impact on the reactions, in general, and the reaction time as well as the reaction yield, in particular. An excess of boronic monomer (2 equiv. for each reaction site) speeds up the reaction and increases the yield. Therefore, we assigned the amount of boronic acid to 2 equiv. for each reactive position. It is noted that the excess boronic monomers do not interfere with the purification. Any residual boronic monomers are effectively absorbed by the silica gel and alumina, and, thus, finally removed from the reaction medium through simple filtration. The optimum yields of **Re1** and **Re2** using a methyliminodiacetic boronic acid ester (MIDA) reactant are 98% and 96%, respectively. Such values ease the purification process. Notable is the scale for building blocks with MIDA similar to that of boronic acid or boronic ester. To extend our material library, several building blocks with boronic acid or boronic ester groups were selected for synthesis. After optimization, the **Re1** yield using boronic acid was boosted to 95%, similar to that of the MIDA reactant (Supplementary Table 1. and Supplementary Figure 2).

Recrystallization is an effective way to purify organic semiconductors, especially those with short or no alkyl chains at all. Compared to chromatographic methods, recrystallization is a better way to purify a large number of compounds in parallel. It is also much easier to obtain purified products by simple filtration. Recrystallization is rarely employed in high-throughput synthesis, especially in cases where molecules with large structural variations are expected[14, 18, 21-22]. This is due to the time-consuming process of finding an optimal solvent. The solvent-antisolvent method and the solvent ratio are hard to transfer to molecules with different structures. Here we adapted a high-throughput platform that integrated solution distribution, heating, stirring, and cooling to screen the optimal solvent composition for recrystallization.

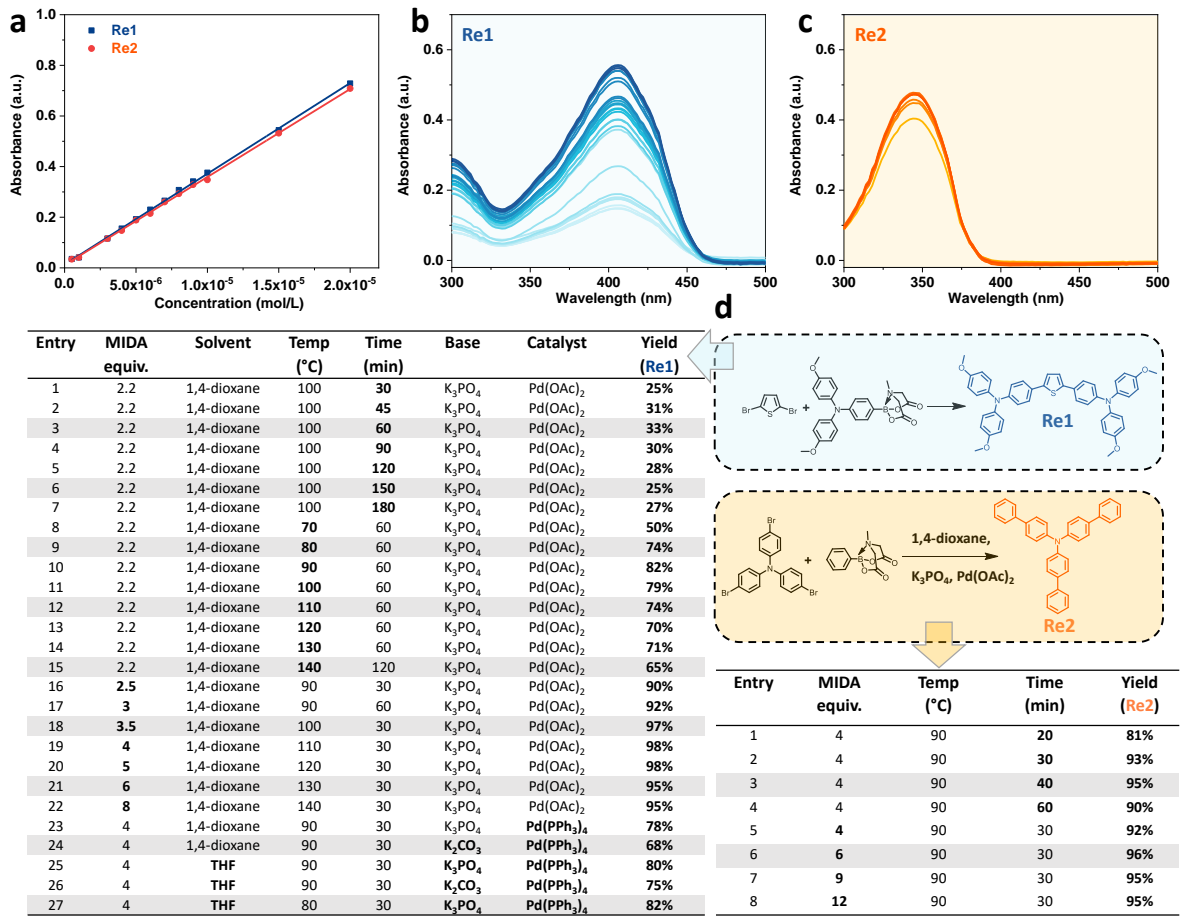

**Figure 2. Optimization of reaction conditions based on reference materials. a**, absorbance-concentration curve for **Re1** and **Re2** in the range from $5 \times 10^{-7}$ M to $2 \times 10^{-5}$ M. **b** and **c**, UV-vis absorption of the diluted reaction solution at different conditions. **d**, chemical structures and synthetic routes of **Re1** and **Re2**.

Nine good solvents (chlorobenzene, toluene, 1,4-dioxane, tetrahydrofuran (THF), chloroform ($CHCl_3$), ethyl acetate, *N,N*-dimethylformamide (DMF), acetonitrile) and three poor solvents (methanol (MeOH), hexane and cyclohexane) were selected for the preliminary solvent screening. Among them, THF and $CHCl_3$ were chosen as good solvents, and MeOH and hexane as anti-solvents. The ratio of good solvent to poor solvent varied from 3:1 to 1:3. The detailed solvent ratio and the corresponding yields are collected in Supplementary Table 2-3, Supplementary Figure 3-4. The whole process was also recorded in Video 1. Supplementary Figure 3 and Supplementary Figure 4 clearly show that the mix-solvent of THF:Hexane (1:2) gives the highest isolated yield of 81% for **Re1** and 92% for **Re2**. For the mixed solvent, when the concentration of anti-solvent is more than 75%, that is, 1:3, it is difficult to dissolve the materials. Therefore, in the subsequent high-throughput purification of the 125 molecules, we took THF:Hexane (1:2) as the preferred solvent for recrystallizing molecules with triphenylamine at the periphery and THF:MeOH (1:2) for those with either triphenylamine in the core or without triphenylamine.

**High-throughput syntheses and purification.** Having optimized the entire process chain for **Re1** and **Re2**, we targeted the semi-automated synthesis and purification of a wide range of conjugated small molecules. To this end, 45 bis, tris, or tetrafunctional bromides and 19 boronic acid derivatives (Figure 3a and 3b, respectively) were chosen from thousands of commercial monomers. A general selection criterion is to diversify the synthesized molecules, including their conjugation length, heteroatom species, solubility, band-gap, mobility, dipole, energy, etc. We named the products according to the combinations of the serial numbers of two reactants (Supplementary Figure 5). The detailed chemical structures and their names are summarized in Supplementary Figure 6. Similar to the semi-automated **Re1** and **Re2** syntheses, all of the generated products were purified using filtration and recrystallization.

During the synthesis process, we did not directly combine monomers **A** and **B**. Instead, we first combined **A1-A30** with **B1-B2** to obtain 54 molecules. Among them, six **A** monomers were randomly removed during the synthesis of **AxB2**. **B1**, 4-methoxytriphenylaniline, is a commonly used functional group in organic semiconductors, especially in perovskite solar cells, due to the excellent electron-donating properties and solubility. On the other hand, **B2**, benzo[b]thiophene, has a distinctly different structure from B1, which aligns with our monomer selection principle of diversifying molecules. Based on the material properties of the first 54 molecules, and by incorporating chemical intuition, we predicted new molecular structures and synthesized them, resulting in samples numbered 54-101. In the second iteration, machine learning was applied to build the model and predict new molecules (molecules 102-125). This will be reported in a separate paper. As a result, starting from molecule 102, the sequence numbers of monomers are not consecutive and very large.

Following optimization of the reaction conditions, only a few by-products were present in the reaction mixture. However, the large variation in solubility and crystallinity due to the drastic differences in molecular structure render purification very challenging. After a single purification cycle, some products needed further purification due to low purity or sample weight. At this point, we created purification encodings (Figure 3c), which comprised sample property encoding and further action encoding. Taking the sample property encoding as an example, the width of the encoding corresponded to the number of sample properties while bits represented properties that were present in a material, similarly to one-hot encoding[23]. The purification encodings, on the other hand, can be adjusted according to the actual requirements. As a matter of fact, this is beneficial for an upgrade to a fully automatic purification. In the current context, attempts to optimize the isolated yields were kept moderately. Instead, we focus on obtaining products with suitable purity and sample weight as quickly as possible, which is typical in most medical chemistry screening campaigns[22].

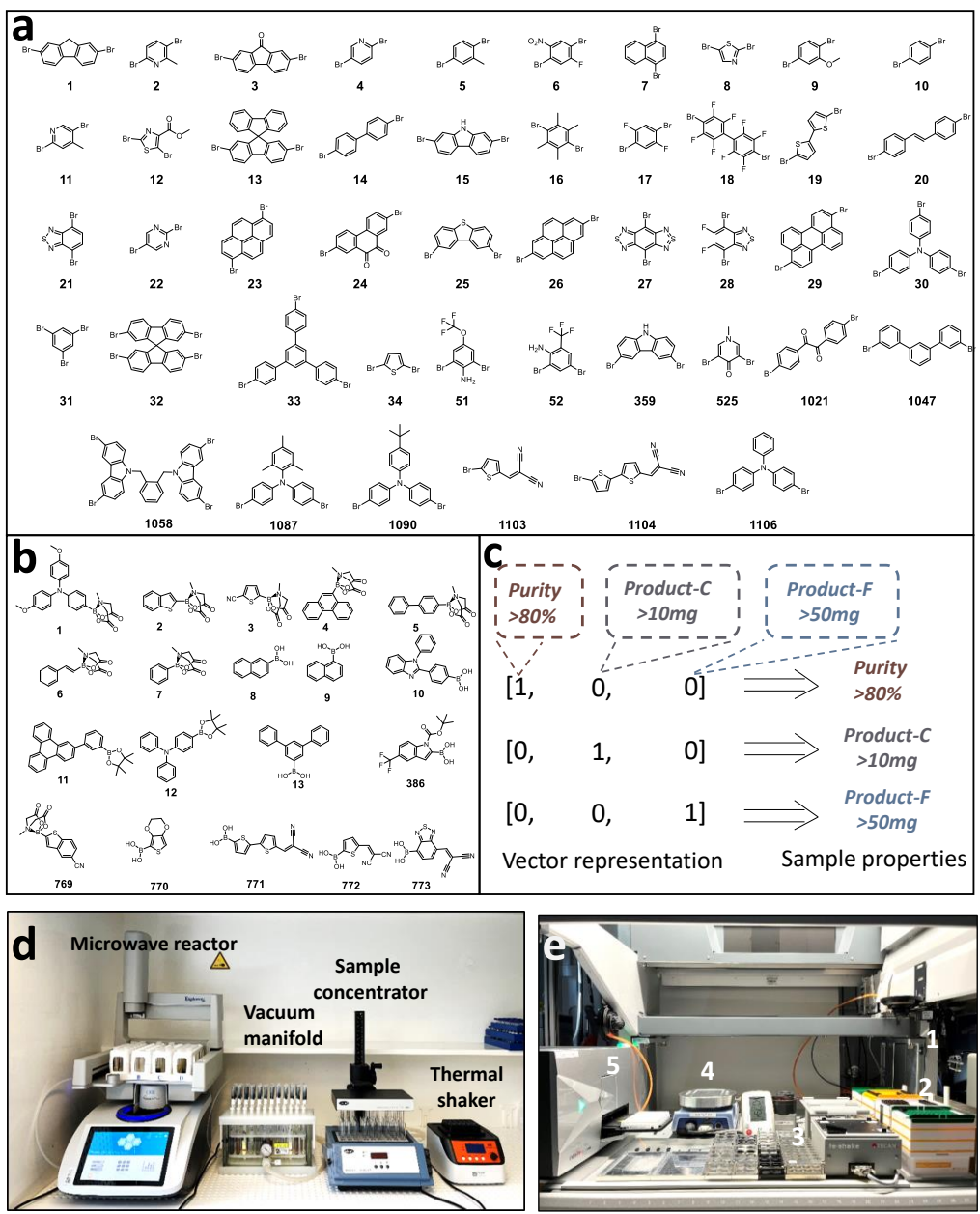

**Figure 3. High-throughput synthesis and purification. a**, Brominated building blocks (Monomer A); **b**, Boronic acid building blocks (Monomer B) used in the coupling reactions. **c,** encoding sample properties. **d**, photograph of high-throughput setups. **e**, TECAN system: (1) robot arm with four pipettes; (2) pipet tips; (3) stock solutions; (4) hot plate; (5) microplate reader.

Therefore, we focused on the following properties: purity, weight of crystals recrystallized from filtered solution (Product-C), and weight of products obtained from the filter cake (Product-F). This was meant to determine how to proceed from here on. For example, [0,1,0] represents a product that features low purity (<80%) after purification and moderate weight (>10mg). But it is free of any filter cake. In other words, the product has high solubility but low crystallinity. Then, the product was recrystallized again, but with more anti-solvent. The corresponding encoding is [0,0,1,0,0]. All sample properties, together with an analysis of the corresponding factors, are listed in Supplementary Table 4. Photographs of the purification process are shown in Supplementary Figure 7-9.

**HT characterizations**. All samples were characterized by $^1$HNMR, UV-Vis absorption, PL spectroscopy, and mobility. Some of the products were also characterized with matrix-assisted laser desorption/ ionization time of flight mass spectrometry (MALDI-TOF MS) and CV to confirm the purity of products and the value of HOMO based on density functional theory (DFT) calculation, respectively. Figure 4 summarizes all products' sample weight, yield, and purity. Among 125 reactions, we obtained 111 samples in more than 10 mg and 90 samples with a purity of more than 70%, sufficient to meet the requirements for functional discovery assays. In Figure 4b, several molecule solutions were diluted owing to the too high UV-vis absorbance (>1) or PL value (exceeding the machine detection threshold). The PL intensity-concentration curve of those molecules can be seen in Supplementary Figure 12.

One advantage of our HT synthesis is being able to generate a big material library quickly. Combining our HT characterization equipment, including the TECAN system with absorption and PL spectroscopy[24-25], next to an automated spin coating station, we gathered all relevant data in a matter of several days. Figure 4 illustrates that the variation in absorption, PL, mobility, and energy levels are rather large. Such an asset is tremendous as it allows selecting from any library depending on the specific needs. For example, materials with suitable mobility and energy levels could be used as transporting materials for semiconductor devices. In addition, building a material library opens the ways and means to discover materials with unexpected properties. Taking, for example, **A31B3** with a simple chemical structure. It exhibits an excitation wavelength-dependent emission (Ex-De) (Supplementary Figure 13), which violates Kasha's rule[26]. In other words, in **A31B3,** photons are emitted from higher-lying excited states, not just the lowest excited state. Most reported EX-De materials are, however, either nanoparticles or metal complexes.[27-29] Our material library provides a compelling strategy to construct purely organic EX-De materials. Of great importance is the fact that all data are obtained under the same conditions. This fact facilitates widely applicable structure-property relationships, which have been constructed in a separate article, where the products were employed as hole-transporting materials in perovskite solar cells.

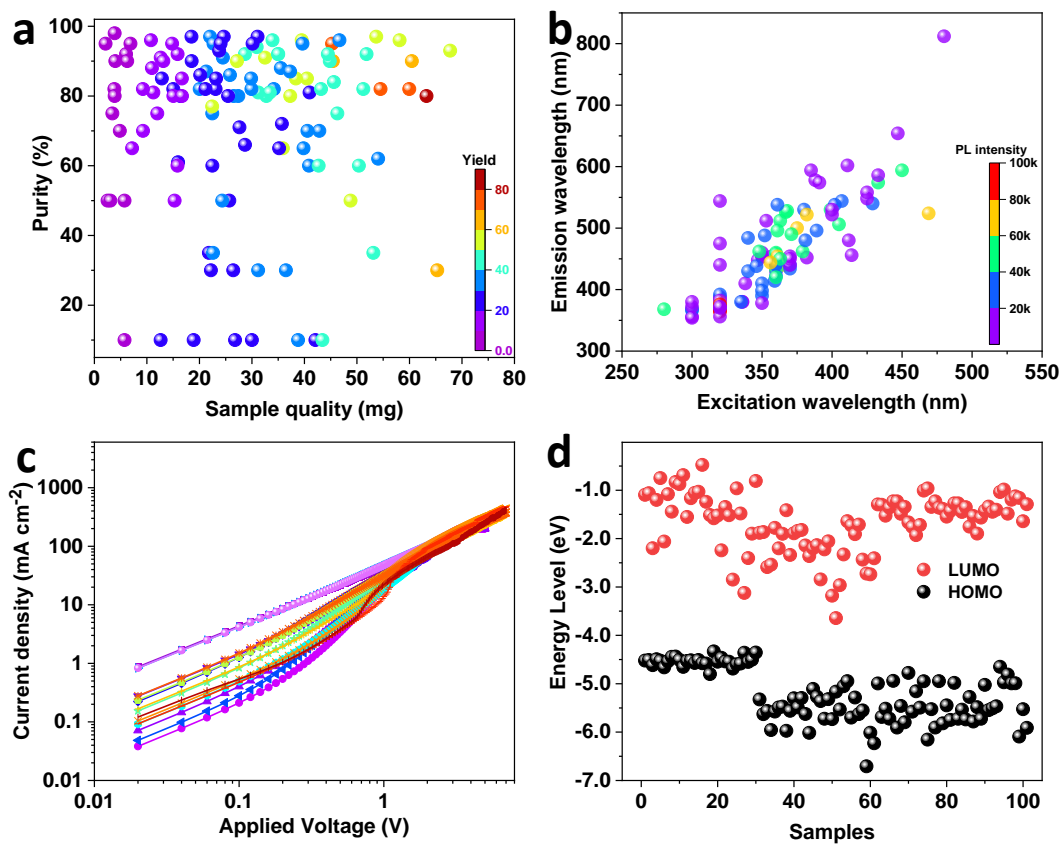

**Figure 4. Properties of the synthesized compounds. a**, sample weight, yield, and purity. **b**, UV-vis absorption and PL. **c**, hole mobility of the partial sample (film); **d**, HOMO and LUMO based on DFT calculation.

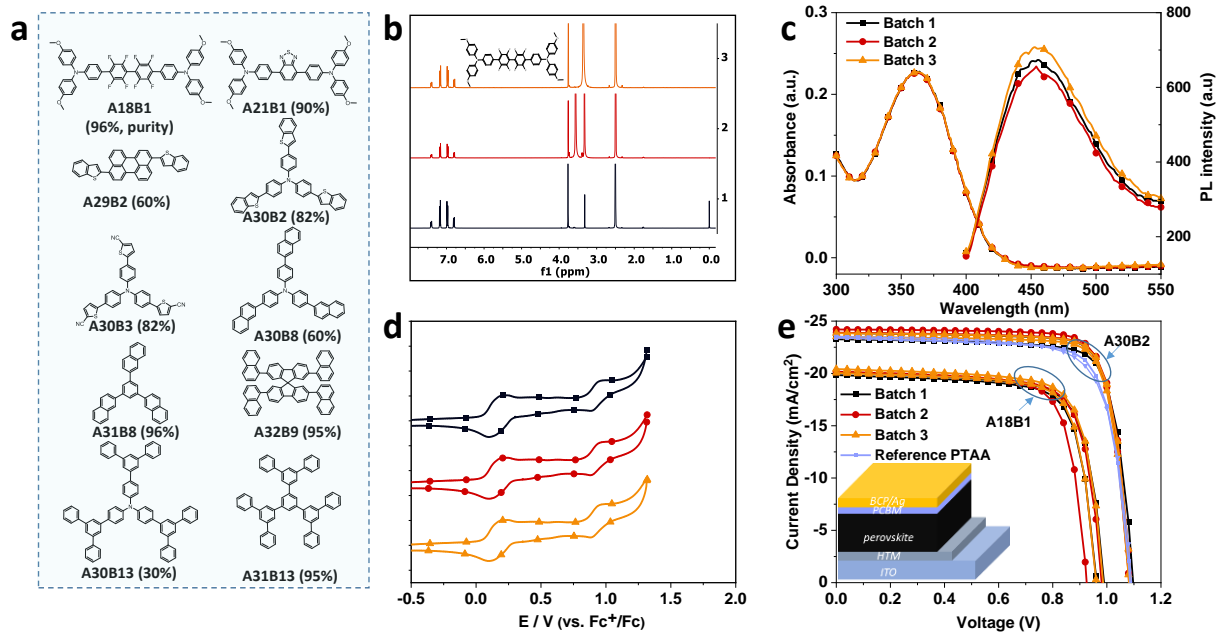

**Figure 5. Characterization of batch-to-batch repeatability**. **a**, Chemical structures of 10 molecules with different purity selected for batch-to-batch repeatability characterization. Characterization of **A18B1** in 3 batches. **b**, $^1$H NMR. **c**, UV-vis and PL spectra (in DMF). **d**, cyclic voltammetry. **e**, performance of **A18B1** and **A30B2** in perovskite solar cells.

**Batch-to-batch reproducibility of the platform**. For our synthesis method, some of the products are inevitably of low purity (< 80%). They may contain reaction products, self-coupling products, partial substitutions, or other by-products. Therefore, batch-to-batch reproducibility is critical. Reproducibility was verified by selecting and repeatedly synthesizing 10 representative products (Figure 5a). Emphasis was placed on different purities before comparing their properties. We considered **A18B1** as an example to prove the repeatability. Figure5b-e shows the spectra of three different batches of **A18B1**. $^1$H NMR shows almost the same peaks in the different batches. Only subtle differences were noted with respect to the solvent peak (3.5 ppm, MeOH) in the second batch. MALDI-TOF MS and thin layer chromatography (TLC) further confirm the same purity and components in each batch (Supplementary Figure 17-38). Even for samples of low purity (<80%), TLCs reveal the same spots regardless of the respective batch. In other words, each batch contains the same by-products or impurities. When organic semiconductors are applied in devices, trace impurities may cause significant changes in mobility, energy levels, PL quenching, etc. As such, the device performance might vary largely. Those changes are believed to be the main reason for semiconductor device degradation. Importantly, we are talking about impurities that might not be detectable by NMR or MALDI-TOF MS. Therefore, those materials in different batches were also characterized by absorption, PL, CV, and mobility, which show negligible alterations (Figure 5c-e). Data for the other nine materials are collected in Supplementary Figure 17-42. In order to understand the electronic quality and the reproducibility of the molecules with respect to electronic quality, we have decided to manually test various batches of two selected molecules as hole transporting layer in perovskite solar cells. (Figure 5e). Among them, the devices based on **A18B1** have moderate performance, while **A30B2** shows excellent performance, with an open-circuit voltage (*Voc*) of 1.08 V, a short-circuit current density (*Jsc*) of 24.19 mA/cm$^2$, and a fill factor (FF) of 0.81, and PCE of 21.04%, even surpassing the commonly used PTAA (Poly(triaryl amine)). What's more noteworthy is their repeatability. Three batches of **A30B2** have almost consistent device performance, including $J_{sc}$, $V_{oc}$, FF, while the performance of **A18B1** devices only fluctuated slightly in the second batch, which may be caused by the remaining small amount of solvent (Figure 5b). This greatly demonstrates the usability of our materials and the reliability of our HT synthesis platform.

The ideal situation is to develop a complete workflow that will couple automatic synthesis with automatic device processing that provides feedback loops. Such feedback loops will recommend new structures based on device results, and allow for a closed optimization of the material to the target criteria of the solar cell. However, such a workflow is far beyond the current paper. Here, we only developed the first part of the workflow, HT synthesis, which is the foundation for the entire process to proceed.

## CONCLUSION

In summary, we present a highly reliable semi-automatic platform to synthesize and characterize solution-processable small-molecule semiconductors. Within several weeks, it can build a material library containing 125 conjugated small molecules and establish their optical and electric properties, encompassing absorption, PL, film mobility, and electrochemistry property, both in theory and experiment. This will accelerate not only the establishment of structure-property relationships based on big data obtained under the same conditions but also the discovery of novel molecules. As the material library becomes available to scientists in various fields, more quantitative structure-property relationships and materials with unexpected properties will be constructed and discovered.

## ASSOCIATED CONTENT

**Supporting Information**

Synthesis and characterization details; computational details; all synthesized chemical structures; $^1$H NMR, MALDI-TOF MS, cyclic voltammetry, UV-vis data; figures of high throughput setups.

**Author Contributions**

The manuscript was written through contributions of all authors. All authors have given approval to the final version of the manuscript. J.W., J.H. and C.J.B. conceived the idea and supervised the project. J.W. performed the high-throughput synthesis. J.Y. characterized the UV-vis and PL. P.R., L.T. and P.F. worked on simulations. L.L and O.K. performed the electrochemical test, J.S.R.O. carried out the MALDI-TOF MS, Y.W. performed on the mobility test. J.W., J.Y., M.H., A.B., Y.Z., Z.X., and J.L. analyzed the data. M.E.P.O. discussed the building blocks selection criteria. J.W. wrote the manuscript.

**Notes**

The authors declare no competing financial interest.

## ACKNOWLEDGMENTS

J.W. and J.L. acknowledge the financial support from the Sino-German Postdoc Scholarship Program (CSC-DAAD). J.W. thanks G.Song for the PL data discussion. C.J.B. gratefully acknowledges financial support through the "Aufbruch Bayern" initiative of the state of Bavaria (EnCN and "Solar Factory of the Future"), the Bavarian Initiative "Solar Technologies go Hybrid" (SolTech), and the German Research Foundation (DFG) SFB 953–No. 182849149. J.W., J.H., Y.Z. and C.J.B. gratefully acknowledge the grants "ELF-PV-Design and development of solution processed functional materials for the next generations of PV technologies" (No. 44-6521a/20/4) by the Bavarian State Government. J.Z., Z.X. acknowledge financial support from the China Scholarship Council (CSC). Y.Z. acknowledges the Alexander von Humboldt Foundation for supporting his scientific research during the postdoctoral period (Grant No. 1199604). J.S.R.O gratefully acknowledge German Academic Exchange Service (DAAD) for its program Research Grants – One-Year Grants for Doctoral Candidates. Calculations were performed on the HoreKa supercomputer funded by the Ministry of Science, Research and the Arts Baden-Württemberg. P.R., L.T. and P.F. acknowledge support by the state of Baden-Württemberg through bwHPC. J.L. gratefully acknowledge financial support from the National Natural Science Foundation of China (Grant No. 62104031) and the Technical Field Funds of 173 Project (Grant No. 24JJ210663A). J.W. acknowledge Fluorochem for providing initial monomer libraries.**REFERENCES**
(1) Duan, L.; Qiao, J.; Sun, Y.; Qiu, Y., Strategies to Design Bipolar Small Molecules for OLEDs: Donor-Acceptor Structure and Non-Donor-Acceptor Structure. *Adv. Mater.* **2011,** *23* (9), 1137-1144.
(2) Lin, Y.; Li, Y.; Zhan, X., Small molecule semiconductors for high-efficiency organic photovoltaics. *Chem. Soc. Rev.* **2012,** *41* (11), 4245-4272.
(3) Chen, Y.; Wan, X.; Long, G., High Performance Photovoltaic Applications Using Solution-Processed Small Molecules. *Acc. Chem. Res.* **2013,** *46* (11), 2645-2655.
(4) Rodríguez-Seco, C.; Cabau, L.; Vidal-Ferran, A.; Palomares, E., Advances in the Synthesis of Small Molecules as Hole Transport Materials for Lead Halide Perovskite Solar Cells. *Acc. Chem. Res.* **2018,** *51* (4), 869-880.
(5) Zhang, L.; Zhou, X.; Liu, C.; Wang, X.; Xu, B., A Review on Solution-Processable Dopant-Free Small Molecules as Hole-Transporting Materials for Efficient Perovskite Solar Cells. *Small Methods* **2020,** *4* (9), 2000254.
(6) Mas-Torrent, M.; Rovira, C., Novel small molecules for organic field-effect transistors: towards processability and high performance. *Chem. Soc. Rev.* **2008,** *37* (4), 827-838.
(7) Miao, Q.; Pu, K., Organic Semiconducting Agents for Deep-Tissue Molecular Imaging: Second Near-Infrared Fluorescence, Self-Luminescence, and Photoacoustics. *Adv. Mater.* **2018,** *30* (49), 1801778.
(8) Brabec, C. J.; Cravino, A.; Meissner, D.; Sariciftci, N. S.; Fromherz, T.; Rispens, M. T.; Sanchez, L.; Hummelen, J. C., Origin of the Open Circuit Voltage of Plastic Solar Cells. *Adv. Funct. Mater.* **2001,** *11* (5), 374-380.
(9) Scharber, M. C.; Mühlbacher, D.; Koppe, M.; Denk, P.; Waldauf, C.; Heeger, A. J.; Brabec, C. J., Design Rules for Donors in Bulk-Heterojunction Solar Cells—Towards 10 % Energy-Conversion Efficiency. *Adv. Mater.* **2006,** *18* (6), 789-794.
(10) Shirakawa, H.; Louis, E. J.; MacDiarmid, A. G.; Chiang, C. K.; Heeger, A. J., Synthesis of electrically conducting organic polymers: halogen derivatives of polyacetylene, (CH). *J. Chem. Soc., Chem. Commun.* **1977,** (16), 578-580.
(11) Levi, B. G., Nobel Prize in Chemistry Salutes the Discovery of Conducting Polymers. *Phys. Today* **2000,** *53* (12), 19-22.
(12) Cankařová, N.; Schütznerová, E.; Krchňák, V., Traceless Solid-Phase Organic Synthesis. *Chem. Rev.* **2019,** *119* (24), 12089-12207.
(13) Briehn, C. A.; Bäuerle, P., Design and Synthesis of a 256-Membered π-Conjugated Oligomer Library of Regioregular Head-to-Tail Coupled Quater(3-arylthiophene)s. *J. Comb. Chem.* **2002,** *4* (5), 457-469.17